# Ultrasonic Array Characterization in Multiscattering and Attenuating Media Using Pin Targets

Yasin Kumru, and Hayrettin Köymen, Senior *Member, IEEE*

*Abstract*— This paper presents an approach to characterize ultrasonic imaging arrays using pin targets in commercial test phantoms. We used a 128-element phased array transducer operating at 7.5 MHz with a fractional bandwidth of %70. We also used a tissue-mimicking phantom in the measurements. This phantom consists of pin targets with a 50 $\mu m$ diameter. We excited the transducer with pulsed and coded signals. We used Complementary Golay Sequences to code the transmitted signal and Binary Phase Shift Keying for modulation. We characterized the transducer array by using the transfer function, line spread function, range resolution, and beam width in an attenuating and scattering medium. We showed that the pin targets, which are very thin compared to the diffraction-limited focus of the transducer array, are suitable for the transducer characterization under weak reflected signal conditions.

## I. INTRODUCTION

THE transducer is a crucial electromechanical element of an ultrasound system since it generates and detects ultrasonic waves [1]. Most researchers use commercial ultrasound transducers made by third-party vendors, even without knowing the transducer properties [2]. Using a suitable transducer with consistent and predictable performance for a given task and investigating the transducer effect on the ultrasound signal are vital processes [3]. The transducer characterization and calibration are tools to determine the transducer properties.

A commonly used transducer characterization and calibration method employs an additional already calibrated transducer [4]. In this method, the performance of the transducer to be calibrated is compared to this calibrated transducer. However, it is not always possible to find a calibrated transducer for this purpose.

The primary method to characterize and calibrate the transducer is the reciprocity-based calibration method. The classical implementation of the reciprocity-based method is the three-transducer reciprocity calibration method [4]-[8]. This method requires three transducers and three different pitch-catch measurement setups. Each measurement setup consists of two transducers, a transmitter and a receiver, to measure the voltage across the receiver terminals and the current driving the transmitter [9]. These electrical measurements provide the sensitivity of any one of the transducers. However, the three-transducer reciprocity calibration is a relatively complex and time-consuming approach due to the need for three separate measurement setups and delicate realignments between setup changes [10].

It is possible to determine the transducer properties by using a single transducer with a single measurement setup. A commonly used method of this type is the self-reciprocity calibration method [4], [10]-[15], which is very suitable for limited test volume applications. This method employs a pulse-echo measurement with a single transducer calibrated with a perfect reflector. A pulse shorter than the total flight time is transmitted, and the driving current is measured. This pulse impinges on the reflector, and the same transducer receives the reflected signals. The transducer is then switched to open circuit receive mode, and received pulse voltage is measured.

Various approaches utilize different types of excitation signals, such as short pulses [16]-[18], discrete frequency tones [1], [19], and linear frequency sweeps [20] for transducer characterization and calibration. In all these methods, authors assume that the individual array elements are identical. A method for the transducer characterization, suggested in [21], characterizes the individual transducer array elements to better predict the transducer array performance. They performed the characterization for different transducers, including piezoelectric transducers (PZT) and capacitive micromachined ultrasonic transducers (CMUT). They showed that the individual element characterization provides a complete transducer evaluation and improves the measurement accuracy. Another study, [22], focused on the transducer functionality to make a characterization. They experimentally investigated the effect of the transducer defect levels on image quality. They made individual element characterization and proposed an acceptance criterion based on the transducer functionality.

Different approaches were also suggested in the literature to make a transducer characterization. In [23], the authors used photoacoustic imaging, which utilizes laser excitation and ultrasound acquisition. They obtained the receive impulse responses of PZT and CMUTs operating at 10 MHz with phantom experiments. In [24], the transducer characterization for high-frequency ultrasound (> 20 MHz) applications is

---

This work was supported by the Scientific and Technological Research Council of Turkey (TUBITAK) under project grant 119E509.

Y. Kumru and H. Köymen are with the Department of Electrical and Electronics Engineering, Bilkent University, Ankara, Turkey (e-mail: yasin@ee.bilkent.edu.tr).

investigated. They characterized a single-element transducer, linear array, and annular array operating at 40 MHz. They made the transducer characterization in terms of 3-D resolution using different sizes of anechoic-sphere phantom structures.

In this study, we used pin targets in a commercial test phantom to characterize the ultrasonic imaging transducers. We used both pulsed and coded excitations for data acquisition. We made the transducer characterization in an attenuating and scattering medium. The rest of this paper is structured as follows. Section II gives the method used in this study. Section III presents the transducer characterization approach by using pin targets together with the experimental results.

## II. METHOD

### A. Measurement Set-up

We collected the data using an ultrasound research scanner. It is called Digital Phased Array System (DiPhAS, Fraunhofer IBMT, Frankfurt, Germany), and shown in Fig. 1. The receive phase length was 95 $\mu s$. We sampled the recorded raw data at 80 MHz. We used a phased array transducer (Fraunhofer IBMT, Frankfurt, Germany) operating at 7.5 MHz center frequency with a fractional bandwidth of 70%. There are 128 elements in the array, and the element pitch is 0.1 mm. We produced the transmitted signals as pulse width modulated (PWM) signals before applying them to the transducer for transmission. The driving signal amplitude is kept constant.

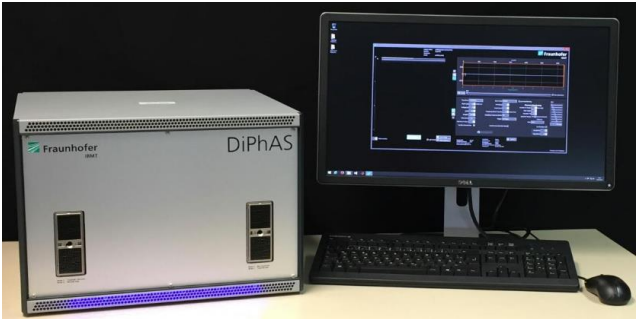

Fig. 1. DiPhAS, Digital Phased Array System, used for data acquisition during the measurements. It is an ultrasound 256-channel research system integrated with a personal computer.

We used a phantom (Model 550, Breast & Small Parts Phantom, ATS Laboratories, Bridgeport, USA) in the measurements. Fig. 2 shows the phantom structure. It is constructed of rubber-based tissue-mimicking material. It consists of monofilament nylon line targets (pin targets) and cylindrical targets of varying sizes and contrasts. The pin targets have a diameter of 50 μm. The attenuation in the phantom is 0.5 dB/cm/MHz. The rubber-based tissue-mimicking material has a sound velocity of 1450 m/s ± %1 at 23ºC. We recorded the phantom and ambient temperatures during the measurements. We positioned the transducer array over the pin targets, and we kept the transducer array acoustically in contact with the phantom surface.

The total dynamic range for programmable gain is 45 dB in DiPhAS. We applied 22 dB fixed gain to ensure a sufficient noise signal at each channel. We limited the time-varying gain to 2.3 dB/cm, which also avoid any saturation.

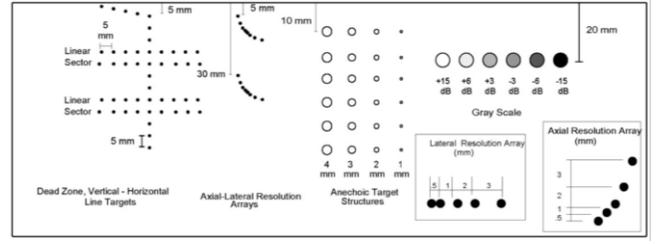

Fig. 2. The structure of the commercial ultrasound phantom used in the measurements. The transducer array is positioned over the pin targets. The pin targets are the ones labeled with "Vertical-Horizontal Line Targets" in the phantom structure.

We also performed measurements in freshwater using the measurement setup shown in Fig. 3(a). The freshwater is almost attenuation-free. The pulse-echo measurements detailed in this section allows us to measure the bandwidth and the two-way response of the transducer for various input pulse waveforms. We also used the data for the design of reference signals, which we used in the correlation receiver. We submerged a highly reflecting material, a steel plate, into the water at approximately 5 cm depth. The thickness of this plate is 15 mm, which is large enough to avoid the interference of the bottom reflection. We excited the mid-element (the 64[th] element) of the transducer. All the elements of the same transducer receive the reflecting echoes. Fig. 3(b) illustrates the photo of this measurement setup, and Fig. 3(c) shows the transmit and receive transducer diagram. The reflected pressure at the $i^{th}$ element surface is approximately given as

$$P_{RX,i}(r;\omega) \approx \left(\frac{P_{TX,i}(\omega)}{\sqrt{r}}e^{-jkr}\right)\Gamma(\omega)\left(\frac{1}{\sqrt{r}}e^{-jkr}\right) \quad (1)$$

where, the subscripts TX and RX represent the transmission and reception. $i$ represents the transducer array element, and it is 64 in this study. The first term on the right-hand side expresses the pressure field at the field point. $\Gamma(\omega)$ is the reflection coefficient of the target. The last term accounts for the propagation from target to the array element. Here, the return path is approximately 10 cm, i.e., $r \approx 5$ cm, and the propagation is cylindrical.

The pressure reflection coefficient at the steel plate surface is taken as unity (impedance mismatch is 1.5 MRayls to 44 MRayls) and the pressure phasor on the element surface during transmission is given as

$$P_{TX,i}(\omega) = H_{TX,i}(\omega)\,V_{TX}(\omega) \quad (2)$$

where $H_{TX,i}(\omega)$ is the forward electromechanical transfer function of the $i^{th}$ transducer element in a rigid baffle, and $V_{TX}(\omega)$ is the Fourier transform of the voltage waveform used for transmission. The received signal is then given by

$$s_{RX,i}(t) = F^{-1}\{H_{RX,i}(\omega)\,P_{RX,i}(r;\omega)\} \quad (3)$$

where $H_{RX,i}(\omega)$ is the backward transfer function of the $i^{th}$ transducer element. $s_{RX,i}(t)$ is measured at the receiver ADC output in units of Least Significant Bit (LSB).

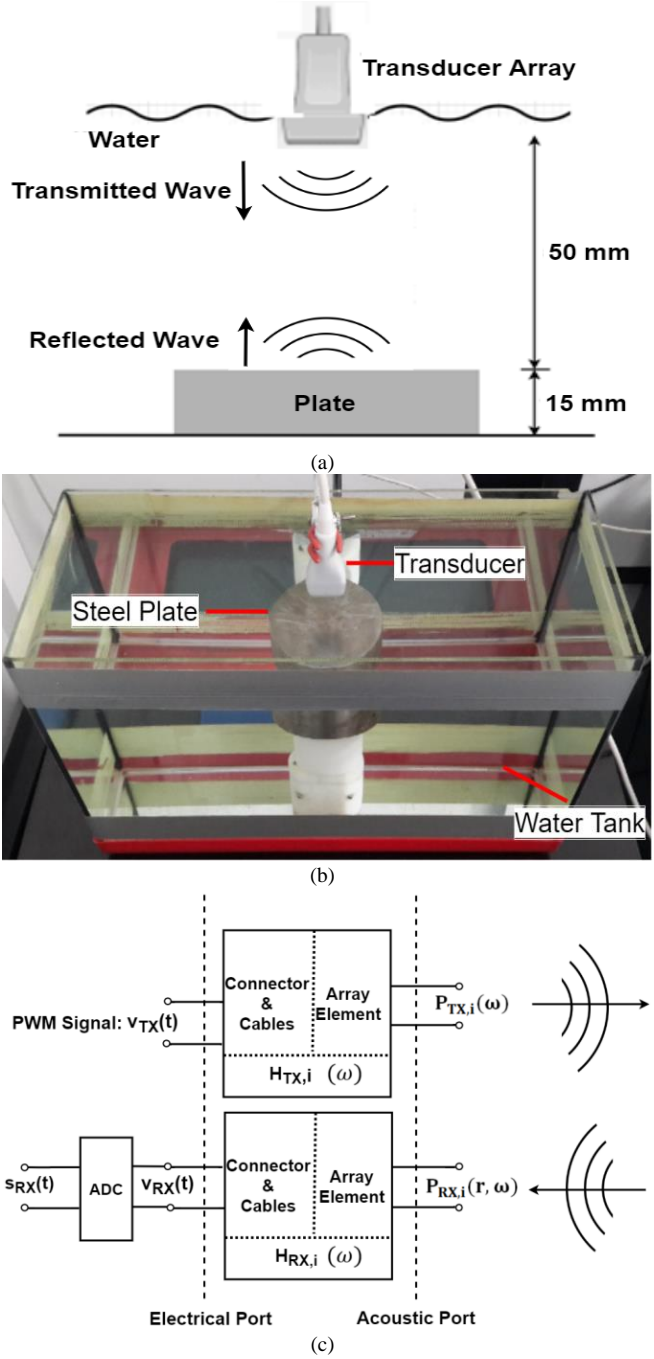

Fig. 3. Freshwater measurement setup. (a) Overview of the measurement setup. (b) A photo of the measurement setup, where the transducer array is fixed and suspended above the steel plate reflector. (c) The transmit and receive diagram of the transducer.

## B. Drive Signals

DiPhAS driver output stages can provide output pulses at a 480 MHz symbol rate. Pulses can have 3-level output voltage, 0 and $\pm V_m$, where $V_m$ is the excitation voltage and can be chosen between 5 V and 75 V. We used 70 V amplitude. It results in an amplitude of +70 V and –70 V so that the peak-to-peak voltage is 140 V.

DiPhAS recommends the PWM signals depicted in Fig. 4. Fig. 4 shows the electrical PWM signals of 2-cycle, 1.5-cycle, and 1-cycle pulses, respectively. These signals are suitable to generate the chip signal for coded transmission. The sampling interval at transmission is 2.083 ns (480 MHz sampling rate). Each half-period pulse contains 12.5 ns 0 V level at the beginning and end of the half-cycle, and 41.7 ns of 70 V amplitude.

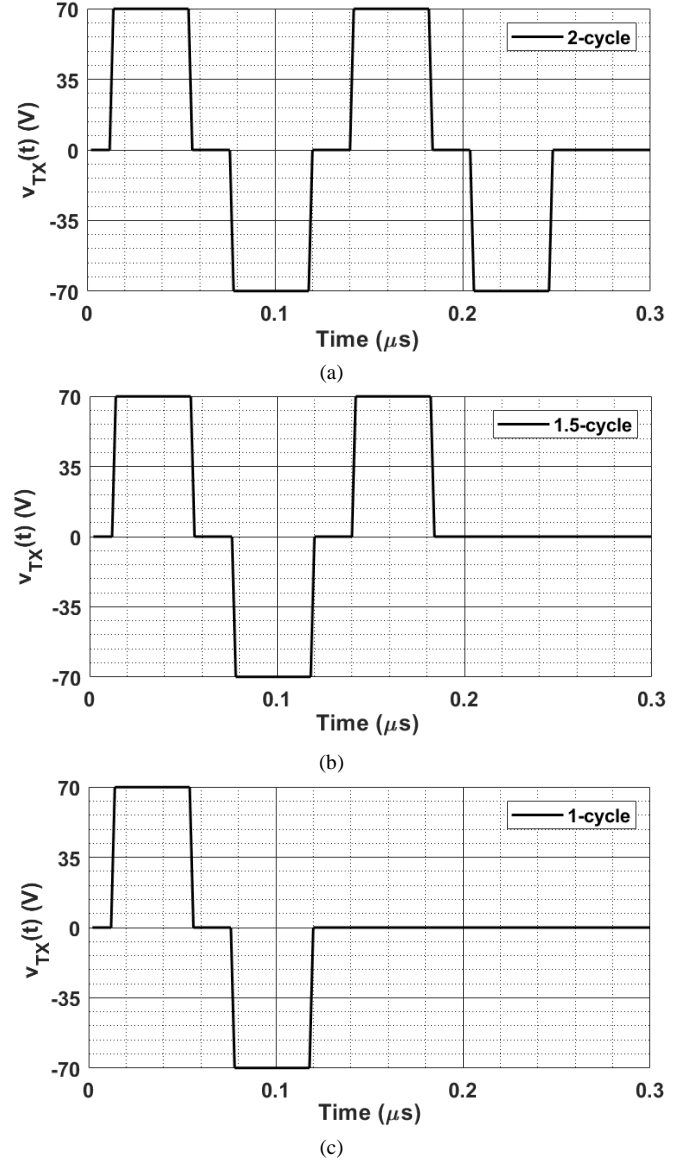

Fig. 4. Electrical PWM drive signals. We used MATLAB Simulink to obtain PWM signals. The PWM signals for (a) 2-cycle pulse, (b) 1.5-cycle pulse (c) 1-cycle pulse.

We also employed a half-cycle signal for imaging purposes. We obtained the insonification signals with the widest bandwidth by using the half-cycle signal. Fig. 5(a) and (b) show the PWM signal for this half-cycle signal and its frequency spectrum, respectively. It is possible to have a wider spectrum if the pulse duration in Fig. 5(a) is shortened. In this case, lower energy signals emerge, and the measurements may suffer from noise problems.

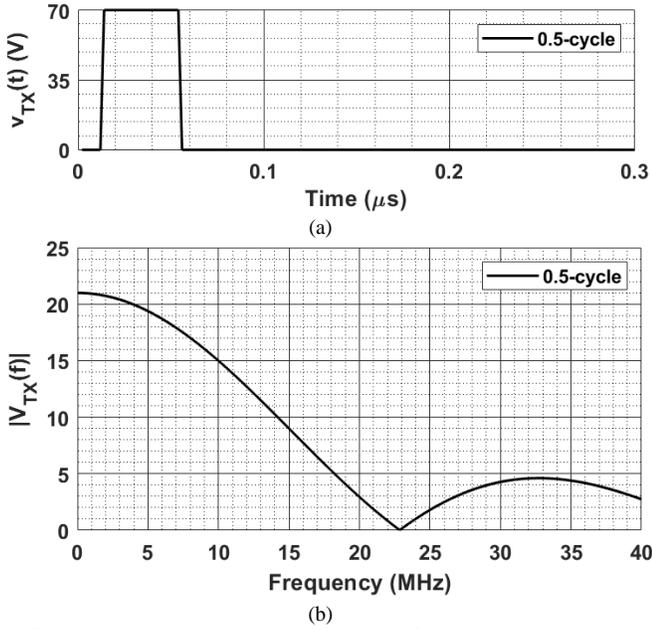

Fig. 5. (a) Electrical PWM drive signal for 0.5-cyle pulse (b) Frequency spectrum of this PWM signal.

*C. Measurement of the two-way electromechanical transfer function of array elements*

We characterized the ultrasound imaging transducer array by measuring the array element transfer function. For this purpose, we transmitted a 0.5-cycle signal. Using the set-up shown in Fig. 3(a), we applied ±70 V pulse to an array element when all other elements were undriven. Then, we measured the reflected signal at the electrical terminals of the same element. Fig. 6 shows the received signal at the 64th element together with its spectrum. This signal includes cumulative effects of the connector, cables, all matching circuits in the transducer assembly. The 3 dB bandwidth of the received signal is less than 5 MHz and is well within the drive signal bandwidth.

The transducer two-way electromechanical transfer function determines the acoustic signal bandwidth. The two-way transfer function, $H_2(\omega)$, is given as

$$H_2(\omega) = H_{RX,i}(\omega) H_{TX,i}(\omega) \quad (4)$$

The measured two-way transfer function, from electrical input to received signal at the same terminal, is shown in Fig. 7(a). The ratio between the frequency responses of the received and the transmitted signals yields the two-way transfer function of that terminal. The 3 dB bandwidth of this element is 4.5 MHz, between 9.92 MHz and 5.37 MHz. This observation agrees with the data provided by the transducer manufacturer given in Fig. 7(b). The measured fractional bandwidth is approximately 67% for a single transducer array element.

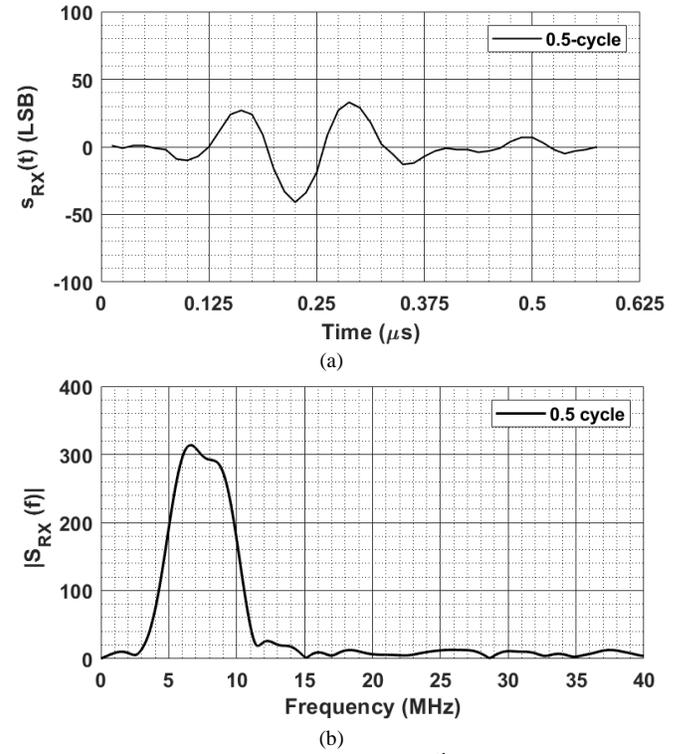

Fig. 6. Fresh-water measurement result. Only the $64^{th}$ element of the transducer array is fired. 0.5-cycle pulse is transmitted. All the elements receive the echo signal. (a) Received signal at the $64^{th}$ element, $s_{RX}(t)$. (b) The frequency spectrum of this received signal, $S_{RX}(f)$.

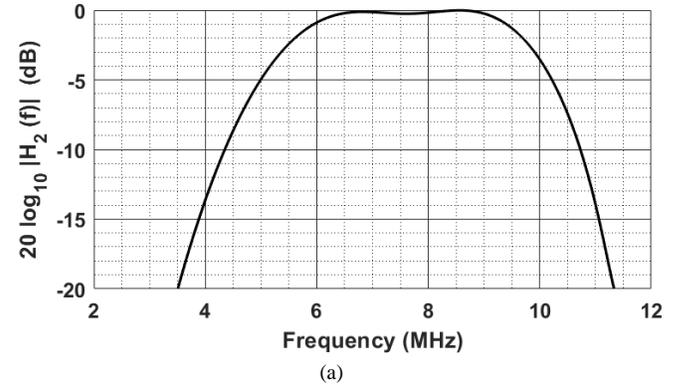

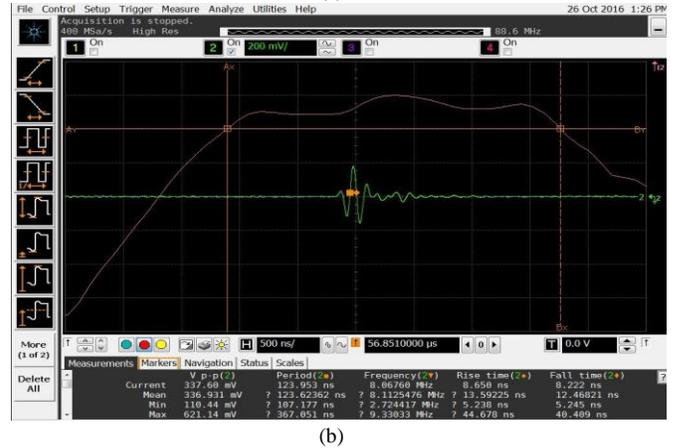

Fig. 7. (a) The measured two-way transfer function of the $64^{th}$ element of the phased array transducer used in the measurements. (b) The data provided by the manufacturer of the phased array transducer (courtesy: Fraunhofer IBMT). The parameters and respective values at the bottom part of the figure can be clearly seen if this figure is enlarged.

## D. The Chip Signals

The transmitted signal distorts, and the transmitted signal energy reduces due to the transducer filtering effect. The energy and the fractional bandwidth (or percent bandwidth) of the received signal quantify this energy loss.

Using the measurement setup shown in Fig. 3, we transmitted pulses with 0.5, 1, 1.5, and 2 cycles at 7.5 MHz center frequency, respectively, from the mid-element of the transducer array. All elements receive the reflecting echoes. The transmitted pulses with different lengths represent the 1-chip of the coded signal. The received signals are substantially different from the electrical driving signals depending on the two-way transducer transfer function. Fig. 8(a) shows the received signals at the $64^{th}$ element of the phased array transducer. 2 and 1.5 cycle signals have sufficient duration to allow the transient response to reach the maximum. The maximum amplitude of the 1-cycle pulse falls off to about 80% and the 0.5-cycle pulse to 40%. The energy in the electrical driving signal of 1-cycle pulse is two times larger compared to 0.5-cycle pulse, as shown in Fig. 4(c) and Fig. 5(a), respectively.

The instantaneous power of a signal is proportional to the squared instantaneous amplitude. This quantity has units of $LSB^2$ in this work. The signal energy over a certain period of time is proportional to the sum of the square of instantaneous amplitude over that period multiplied by receiver sampling interval, $\Delta t$,

$$E_i = \Delta t \sum_{q=1}^{Q} y_i^2(q) \quad (5)$$

and has units of $LSB^2$-s. $\Delta t$ is 12.5 ns in this study. We use the signal energy in this work only for relative comparison and normalization, and we refer to the energy of the signals in units of $LSB^2$-$\Delta t$.

The limitation imposed on the signals by the transducer bandwidth limit is also observable on the maximum peak-to-peak amplitude of the time domain signals. The 2-cycle signal has the maximum peak-to-peak amplitude of 200 LSB, and the 1.5-cycle signal is close to 190 LSB. It is clear that the latter also had enough time for transient response to develop. These two signals have similar maximum amplitude, but the 1.5 cycle signal is wider bandwidth with a shorter duration. 1-cycle signal is lower with 150 LSB, and 0.5 cycle signal has the lowest maximum amplitude of 70 LSB. Short duration signals with large amplitude are important for the performance of diverging wave (DW) applications.

We also estimate the attenuation effect on the signal using the nominal attenuation of the phantom, 0.5 dB/MHz/cm. We first transformed the received signal, $s_{RX}(t)$, into frequency domain and we obtained $S_{RX}(f)$. The attenuated signal, $S_{RXA}(f)$, is then obtained as

$$S_{RXA}(f) = S_{RX}(f) \, e^{-\alpha f (2r)} \quad (6)$$

where $\alpha$ is the attenuation coefficient in Nepers/cm/MHz, $f$ is the frequency in MHz, and $2r$ is the total (round-trip) distance in cm. We then transformed $S_{RXA}(f)$ back to time domain and we obtained $s_{RXA}(t)$. Fig. 8(b) shows $s_{RXA}(t)$ from 4 cm depth for 0.5, 1, 1.5, and 2-cycle excitation, respectively. The amplitudes of all signals are lower by about 20:1 due to the attenuation, although the characteristic signal morphology is generally preserved. The energy of 2-cycle signal is again highest, but the peak amplitude of 1.5-cycle signal is now maximum.

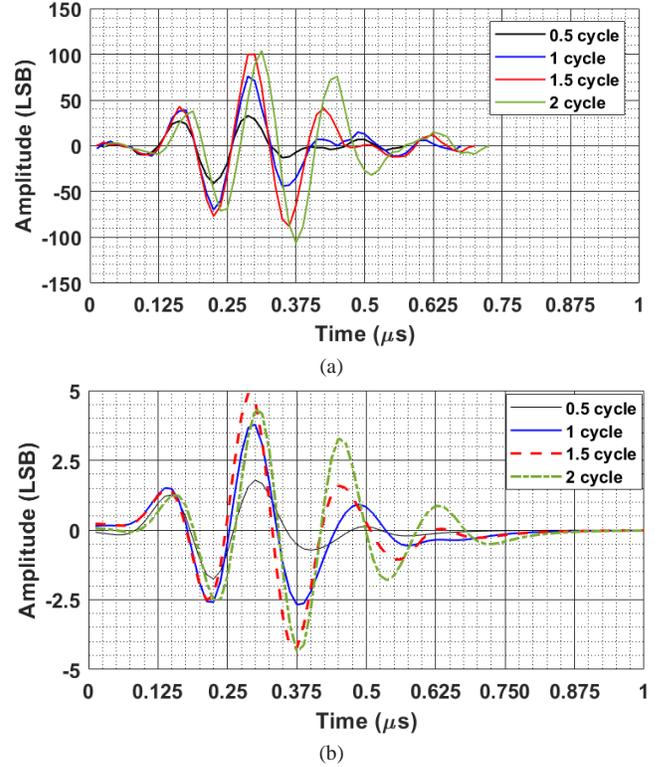

Fig. 8. Pulses with 0.5, 1, 1.5, and 2 cycles. (a) The signals reflected from the steel plate and received by the $64^{th}$ element of the phased array transducer are shown. (b) The attenuation compensation effect is shown. Attenuation compensation with respect to 4 cm depth is applied on the received signals.

We calculated the fractional bandwidth of each received signal. The fractional bandwidth is the full-width at half maximum (FWHM; -3 dB range) of the power spectral density (PSD) divided by its center frequency [25]. The PSD is a measure of the power distribution over frequency, and we computed the PSD by using a periodogram method. For a signal $x_n$ with length N sampled at $f_s$ samples per unit time, the power spectral density estimate is calculated by,

$$P(f) = \frac{1}{N f_s} \left| \sum_{n=0}^{N-1} x_n e^{-j 2\pi \frac{f}{f_s} n} \right|^2 , \quad -\frac{f_s}{2} < f \leq \frac{f_s}{2} \quad (7)$$

Fig. 9 shows the PSDs of the two-way transmitted-and-received signals with different pulse lengths as a function of frequency. The fractional bandwidth computed from PSD estimate is 67% for 0.5-cycle, 62% for 1-cycle, 50% for 1.5-cycle and 39% for 2-cycle pulses. 0.5-cycle signal has the widest bandwidth and undergoes a much larger energy loss compared to the drive signal energy. On the other hand, 2-cycle pulse is a relatively narrow-band signal and suffers the smallest energy loss. The 0.5-cycle pulse has approximately 9 times less energy compared to 2-cycle pulse, whereas the ratio of the energies of the respective drive signals is 4-to-1.

Fig. 9(b) shows the spectrum of the attenuated signals. The spectra are skewed to lower frequencies. The center frequencies are lowered and the bandwidth became narrower as a result of the attenuation.

The attenuated signals provide an insight for the received signal in the phantom. Attenuation causes a significant decrease in the amplitude and a shift of energy to lower frequencies in the spectrum.

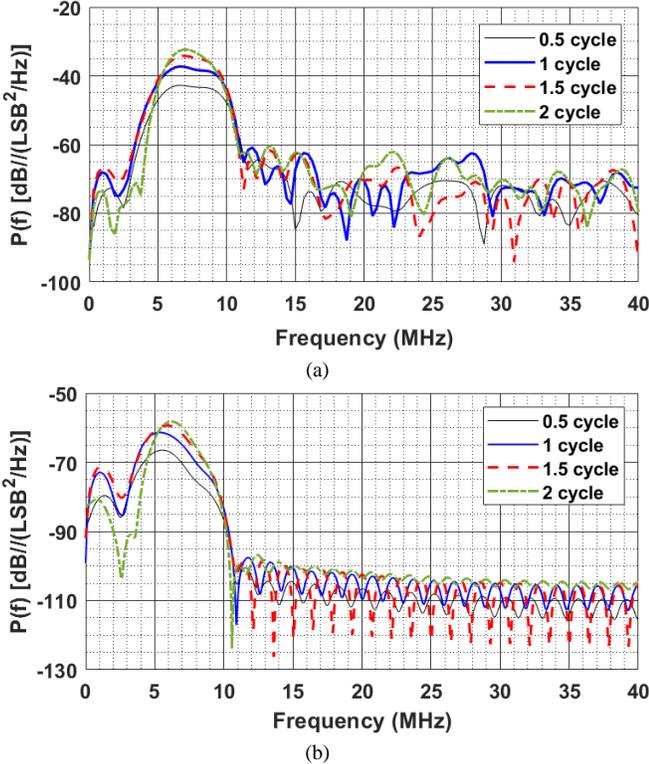

Fig. 9. The PSDs of the signals with different duration as a function of frequency. (a) The PSDs of the two-way transmitted-and-received signals, which are shown in Fig. 8a. (b) The PSDs of the attenuation compensated two-way transmitted-and-received signals, which are shown in Fig. 8b.

*E. Correlation Receiver Output*

We implemented the matched filter as a correlation receiver [26]. A correlation receiver comprises a mixer and an integrator. The inputs to the correlator are the received channel data and the reference signal. The reference signal is the normalized version of the signals depicted in Fig. 8(a). The reference signal is normalized so that it has unit energy.

We performed a measurement in freshwater, as shown in Fig. 3(a). We transmitted 0.5, 1, 1.5, and 2 cycle signals, respectively. We then correlated the received signals at the 64[th] element with the respective reference signals. Fig. 10 shows the correlation receiver output signals. The correlator output for this case is shown in Fig. 10(a), where the received signal is unattenuated. We also applied attenuation on the received signal for 4 cm depth to investigate the attenuation effect on the correlator output. The correlator output for this case is shown in Fig. 10(b).

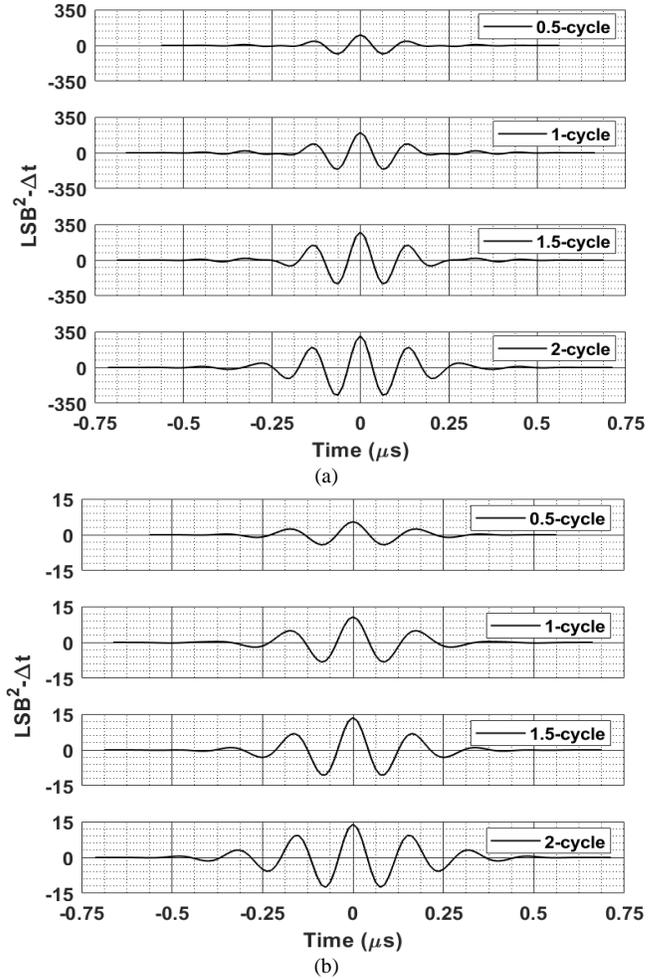

Fig. 10. The correlator output for 0.5, 1, 1.5, and 2 cycle signal excitations in freshwater. The reference signals are the normalized versions of the respective received signals. The reference signals for all excitation signals have unit energy. Hence, the correlator gain remains same for all excitation signals. (a) The correlator output without attenuation (b) Correlator output with attenuation.

Longer signals have higher correlation receiver outputs in both cases. The peak output of the 1-cycle signal is 65% of the peak output of the 2-cycle signal in the un-attenuated case (See Fig. 10(a)). However, this ratio increases to 77% when attenuation is present (See Fig. 10(b)). Similarly, the ratio of 0.5-cycle signal peak output to that of 2-cycle signal increased from 33% to 39% after attenuation. The correlator outputs of signals with wider bandwidth suffer relatively less from propagation in attenuating medium. The wider bandwidth of these signals provides relatively more energy at the lower frequency range, where the attenuation effect is less. 0.5-cycle and 1-cycle signals compare similarly in this respect. Note that the lower cut-off frequency of the transducer response limits the amount of lower-frequency energy.

III. CHARACTERIZATION USING PIN TARGETS

*A. Reflection from pin targets*

In ultrasound imaging, the region of interest is always in the near field of the array. The array elements have sufficiently large dimensions in the transverse direction. Therefore, the propagation of acoustic signals emitted from an array element is predominantly cylindrical in the near field of the array.

The tissue-mimicking phantom contains gratings of pin targets in various spatial configurations, as shown in Fig. 2. The pin targets in the phantom are nylon monofilaments of 50 $\mu m$ diameter. They are cylindrical, and their axis are nominally normal (transverse also) to the plane of wave propagation emitted by the transducer array. Reflection of waves from cylindrical targets is a well-studied subject in acoustics [27], [28].

The pin target radius, $a$, is 1/8 of a wavelength at 7.5 MHz, which makes $ka = 0.785$, where $k$ is the wavenumber. A cylindrical surface of this size reflects the impinging cylindrical waves omnidirectionally but as cylindrical waves. The average $ka$ is even lower for the acoustic pulse in the medium since the energy of the acoustic signals is confined to a lower frequency band as the attenuation becomes effective (See Fig. 9). Line targets, which are very thin compared to the diffraction limited focus of the imaging array, are suitable for acoustical characterization of the transducer array. Considering that such targets are omnidirectional reflectors in the plane of the cylindrical wave without any specular reflection, we developed the measurement procedures to assess acoustical properties of the array. We used these pin targets to measure the line spread function (LSF) [29], range resolution, and beam width of the array in attenuating media.

One particular difficulty in this type of measurement is the inherent low reflected signal levels due to low $ka$ [28]. The low target strength of the pin targets is further aggravated by the fact that they are made of nylon. Nylon is a polymer with density, $\rho$, of $1.14\times10^3$ kg/m$^3$, and sound speed, $c$, of 2620 m/sec, yielding an intrinsic impedance, $\rho c$, of 3 MRayls [30]. This choice of material is excellent for tissue-mimicking phantom in many ways. However, for transducer characterization purposes, a metal wire would yield a much higher echo. For example, the echo amplitude would be three times (10 dB) higher if the pin targets were made of steel wire, which has an intrinsic impedance of 44 MRayls.

We employed matched filter in the receiver to improve the Signal-to-Noise Ratio (SNR) in the measurements. The use of coded signals further improves SNR in measurements for characterization [31], which is particularly instrumental when unfocused transmission is employed, such as plane wave imaging [32]-[34] or diverging wave imaging [35]-[42].

*B. Measurement Technique for two-way Radial and Lateral Resolution*

The resolution in a particular imaging scheme depends on the transducer LSF as well as the imaging scheme. The best resolution is provided by focused transmission schemes *albeit* at the focus region only. In fast imaging schemes, the transmission is unfocused and resolution is achieved by only focusing at reception beamforming.

In order to measure the resolution of the array, we imaged a pin target as described in [31], using any one of the imaging schemes. We calculated the beamformed signal amplitude at every raster point spaced by 25 μm in a 6 mm × 6 mm region centered at the pin target. Fig. 11 depicts the DW image of the center pin at 40 mm depth. We determined the maximum amplitude and plotted the lateral and radial received signal amplitude distribution centered at this maximum.

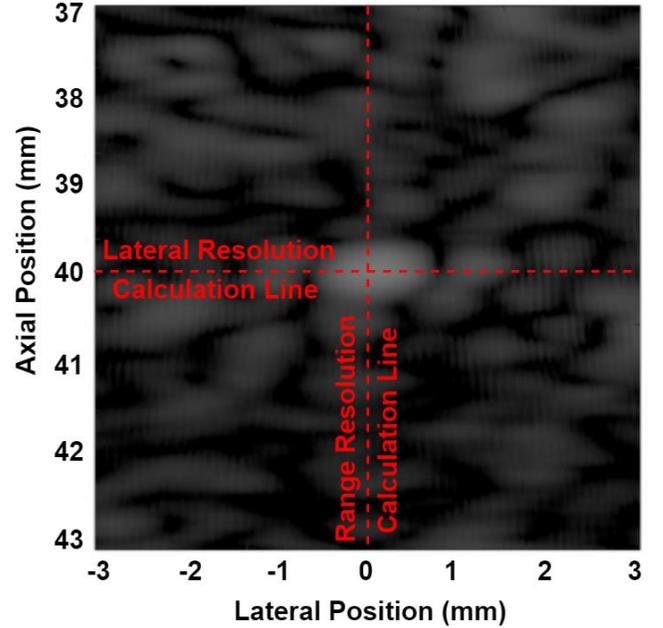

Fig. 11. DW image of center pin at 40 mm depth with 25 $\mu m$ resolution. The line spread function (LSF) is calculated along the lateral resolution calculation line. The range resolution is calculated along the range resolution calculation line.

*C. Resolution Measurements for DW Transmission*

We measured the LSF and range resolution when DWs with 14 mm virtual source distance [31] is employed. The LSF and the range resolution of the center pin target at 40 mm depth for 2-cycle signal is depicted in Fig. 12. We also measured the lateral and range resolution when coded transmission is employed. We used 8-chip Complementary Golay Sequences (CGSs) with 2-cycles/chip for coding and binary phase shift keying (BPSK) for modulation [31]. The CGS(A) and CGS(B) used in the measurements are as follows:

TABLE I
BIPOLAR REPRESENTATION OF THE 8-CHIP CODE SEQUENCES

| Code Length | Sequence Type | Bipolar Representation |
|---|---|---|
| 8 | CGS (A) | {1  1  -1  -1  -1  1  -1  1} |
|   | CGS (B) | {1  1  1  1  -1  1  1  -1} |

The LSF and the range resolution obtained for coded transmission is also shown in Fig. 12 for comparison. It is clear that in both cases the lateral and range resolution are the same, except the measurement is more reliable when coded signal is used. SNR is 12 dB higher in coded transmission [31], which enables more reliable measurements.

The side lobe level in LSF of pulsed measurement is higher. This is probably because of the fact that we did not correct for individual element transmit and receive sensitivity on the signals. The sidelobe levels in coded transmission is rather symmetric and lower. The level is still 2 to 3 dB larger than rectangular aperture equivalent to array size. Lower sidelobe levels are possible with apodization.

The range resolution is shown in Fig. 12(b). Again, the 3-dB range resolution, which is approximately 200 nsec (or 0.31 mm, one and half wavelength at 7.5 MHz), is the same both for 2-cycle signal (1-chip) transmission and 8-chip coded transmission. Extra range lobes are not introduced in coded transmission.

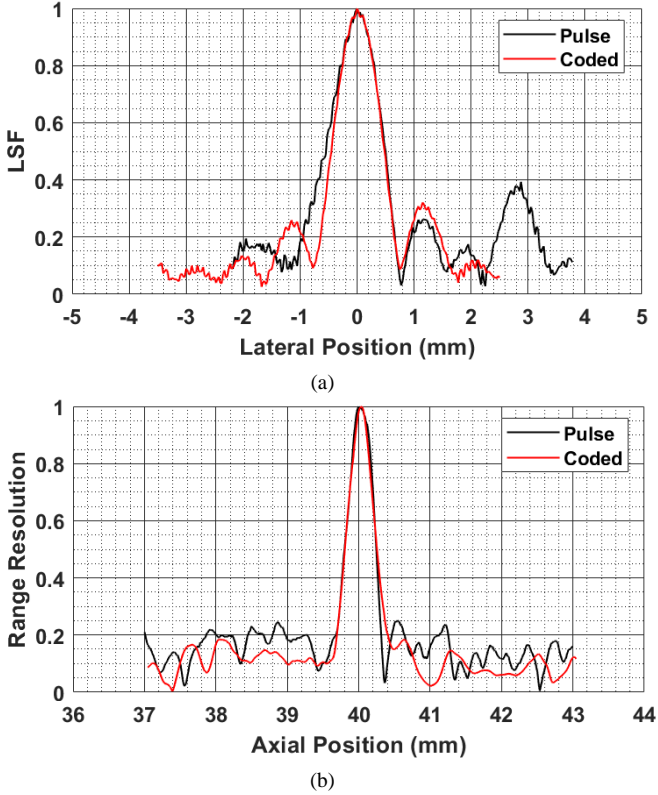

Fig.12. Results for DW transmission. Normalized lateral and range resolution obtained with 2-cycle/chip signal when single chip (1-bit) is transmitted and 8-bit CGS coded signal is transmitted, respectively. (a) LSF, and (b) range resolution.

Fig. 13 shows the resolution of the 8-chip coded signal with 0.5, 1, 1.5, and 2 cycle/chip lengths. The 3 dB LSF width is 0.775, 0.775, 0.700, and 0.675 mm for 0.5, 1, 1.5, and 2-cycle/chip signals, respectively. The 6 dB LSF width is 1.075, 1.05, 1, and 0.925 mm for 0.5, 1, 1.5, and 2-cycle/chip signals, respectively. The lateral sidelobe level is between -10.5 and -13 dB for all chip lengths. This level can be compared with the sidelobe of rectangular aperture, which is -13.5 dB. Sidelobe nulls and maxima are different for each chip length. The variation is consistent with the center frequency difference of the respective spectra given in Fig. 9(b).

The range lobes for wider bandwidth chip signals are better defined.

Considering that the pin target diameter is only 50 μm and the foci are almost a 1 mm wide in lateral direction and about 300 μm in radial direction, pin target is perfectly suitable for this assessment.

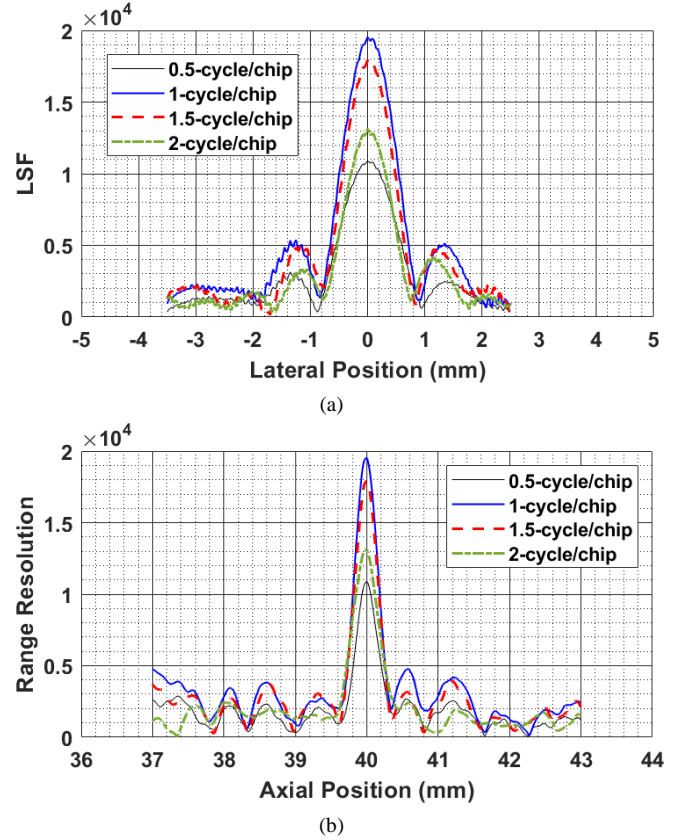

Fig. 13. Results for DW transmission. Lateral and range resolution of 0.5, 1, 1.5, and 2-cycle/chip 8-bit coded transmission signals, (a) LSF and (b) range resolution

### D. Resolution Measurements in Focused Transmission

Measuring the LSF for focused transmission in this method is relatively more involved compared to DW measurements. Complication is due to the necessity of centering the pin target in the focused beam at the focal region. This requires either few transmissions at finely spaced steering angles around the target and pick the most suitable transmission, or use synthetic transmit aperture imaging (STA) data.

### E. Beamwidth Measurements Using Pin Target Grating

We used the grating made of pin targets to determine the beamwidth or pattern of imaging transducer arrays. There are four horizontal gratings of 5 mm spaced pin targets at 20, 25, 40, and 45-mm depth. We utilized the reflections from the pins and determined the lateral distribution of transmitted and received ultrasonic energy. We used focused reception for every field point. This enables us to determine the required phase profile for a given field of view (FOV) for the DW transmission.

We obtained the maximum reflected signal amplitude from any pin target by STA imaging since it enables best focusing in transmission at all field points [31]. These reflected signals are affected by the target position with respect to the array center. The targets further away from the array aperture generate lower echo amplitude due to attenuation (and spreading in DW). Therefore, the relative signal amplitude variation (distribution) obtained by STA imaging along any horizontal pin target grating constitutes a reference for the best distribution of available ultrasonic energy into the sector of interest. If the insonification sector is more acute than required, the reflections from outermost pins are comparatively lower, and SNR is low in regions away from the centerline. It is often possible to use wider spreading phase profile than required, i.e., very short $r_v$, which results in sending a considerable portion of the available energy out of the sector of interest. In this case, the difference between the reflections from the center pin and the outermost pins is less than what is obtained in STA. This section describes how we optimized the delay profile in DW imaging for the insonification of the required imaging sector using the STA reflection signal distribution profile as a reference.

We used DW transmission to insonify the region of interest in sector imaging. DWs can be generated in many ways by applying appropriate delays to array elements [43]. We adjusted the insonification sector by applying delays to the array elements. The waves emanating from the array aperture are part of a cylindrical wave generated by a virtual line source $r_v$ away from the aperture plane [31], [43]. Fig. 14 shows this geometry. The smaller the $r_v$ is, the wider the imaging sector. The geometrical planning of delays results in the insonification of a much wider sector because of diffraction.

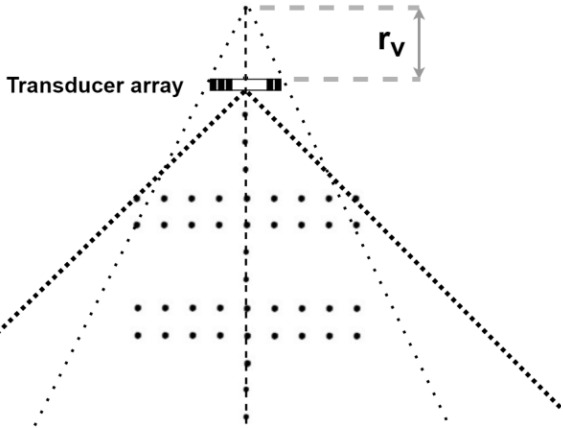

Fig. 14. Geometry for beamwidth determination. We used pin target gratings to determine the beamwidth of the imaging transducer array.

It is also very important to confine the available ultrasonic energy into the region of interest. When the DW is generated using short virtual source distance, the sector is evenly insonified. The drawback is that a significant amount of available energy is transmitted outside of the sector. This results in low reflected signal amplitudes. If large $r_v$ is used, then the energy at the sides of the sector is low. Fig. 15 shows the received signal amplitude reflected from pin target grating at 25 mm depth when DWs are transmitted [31]. Diverging waves are 8-bit coded with 2-cycle/chip. The signal level is highest at the center when $r_v$ is 21 mm and lowest when $r_v$ is 10.5 mm. On the other hand, the transmission using $r_v$ = 10.5 mm yields highest signal level at the outermost pin targets and 21 mm yields lowest. The difference is as large as 14 dB.

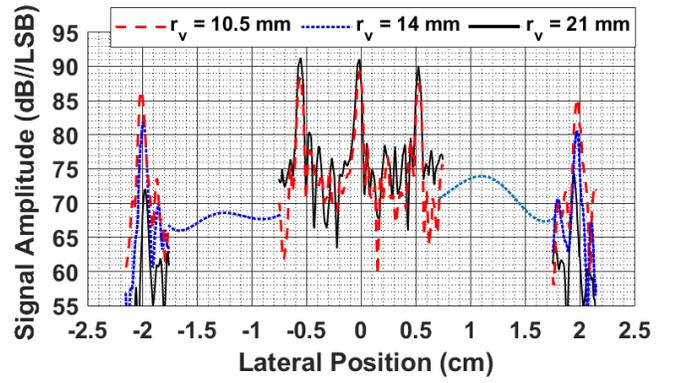

Fig. 15. The reflected signal amplitude of 8-chip coded DW with 2-cycle/chip, as a function of virtual source position, along the horizontally spaced pin targets positioned at 25 mm depth. $r_v$ is increased from 10.5 to 21 mm.

In order to decide on which DW provides adequate insonification across the sector, signal level distribution along the grating obtained by STA provides a reference. Fig. 16 shows the signal amplitude variation of STA (1-chip pulse) and DW with 14 mm virtual source distance along the grating. It is clear that DW has a beam width which provides the same reflection amplitude profile as STA all across the grating. The reflected signal amplitude is only 1 dB lower compared to the one with $r_v$ = 21 mm at the center pin.

Direct comparison of absolute signal amplitudes of two schemes is misleading. There is gain in coded reception at the correlation receiver and there are two transmissions in coded DW transmission.

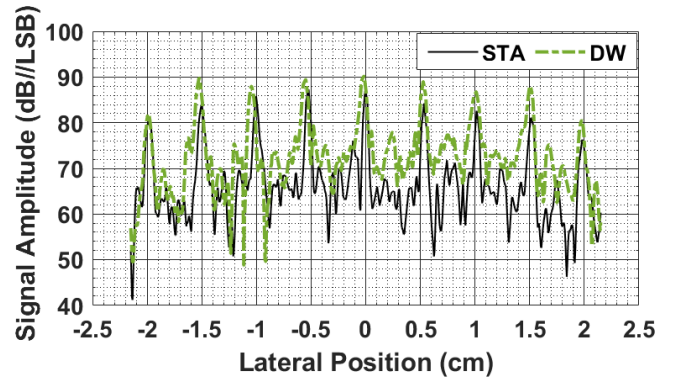

Fig. 16. Signal amplitude for 8-chip coded DW (2-cycle/chip) transmission when $r_v$ =14 mm and STA imaging with 1 chip pulse (2-cycle/chip) along the pin targets located at 25 mm depth.

## IV. CONCLUSION

We characterized our phased array transducer in a multiscattering and attenuating medium. We used the pin targets for transducer characterization, which have a small diameter compared to the diffraction-limited focus of the transducer arrays.

We showed that these pin targets are appropriate for transducer characterization. The reflected wave intensity depends on the pin target material as well as its diameter. The pin targets made of highly reflective material are good choices for transducer characterization since they yield higher reflected wave intensity. In our study, the pin targets are nylon and monofilament, which cause relatively low reflected wave intensity. However, coded transmission compensates this due to the increased SNR and yet yields the same measurement results.

We characterized a 128-element phased array transducer with a 7.5 MHz center frequency. We measured the impulse response of an array element and determined the two-way transfer function. We measured the lateral and range resolution at 40 mm depth in an attenuating medium where the nominal attenuation is 0.5 dB/MHz/cm.

## V. References

[1] K. R. Erikson, "Tone-Burst Testing of Pulse-Echo Transducers," *IEEE Transactions on Sonics and Ultrasonics,* vol. 26, no. 1, pp. 7-13, Jan 1979.

[2] E. K. Sittig, "Transmission Parameters of Thickness-Driven Piezoelectric Transducers Arranged in Multilayer Configurations," *IEEE Transactions on Sonics and Ultrasonics,* vol. 14, no. 4, pp. 167-174, Oct 1967.

[3] J. Mylvaganam, "Characterization of medical piezoelectric ultrasound transducers using pulse echo methods," *MSc Thesis, Norwegian University of Science and Technology,* July 2007.

[4] H. Köymen, "Lecture Notes on Electroacoustic Transduction," Bilkent University, Ankara, 2020, Chapter 9, Acoustic Measurements, https://www.researchgate.net/project/Lecture-Notes-on-Electroacoustic-Transduction, last accessed: May 18, 2021.

[5] R. J. Bobber, "Underwater Electroacoustic Measurements," in *Chapter II: Methods and Theory*, Orlando, Naval Research Laboratory, Underwater Sound Reference Division, 1970.

[6] C. J. Dang, "Electromechanical characterization of ultrasonic NDE system," *Ph.D. thesis, Iowa State University,* 2001.

[7] C. J. Dang, L. W. Schmerr, Jr., and A. Sedov, "Ultrasonic transducer sensitivity and model-based transducer characterization," *Res. Nondestr. Eval.,* vol. 14, p. 203–228, 2002.

[8] H. Hatano, T. Chaya, S. Watanabe, and K. Jinbo, "Reciprocity calibration of impulse responses of acoustic emission transducers," *IEEE Trans. Ultrason., Ferroelect., Freq. Contr.,* vol. 45, no. 5, p. 1221–1228, 1998.

[9] S. Robinson, P. Harris, G. Hayman, J. Ablitt, "Evaluation of uncertainty in the free-field calibration of hydrophones by the three-transducer spherical wave reciprocity method," in *Inter-Noise*, 2016.

[10] A. L. Lopez-Sanchez and L. W. Schmerr, "Determination of an ultrasonic transducer's sensitivity and impedance in a pulse-echo setup," *IEEE Transactions on Ultrasonics, Ferroelectrics, and Frequency Control,* vol. 53, no. 11, pp. 2101-2112, 2006.

[11] E. L. Carstensen, "Self-reciprocity calibration of electroacoustic transducers," *J. Acoust. Soc. Amer.,* vol. 19, no. 6, pp. 961-965, 1947.

[12] R. M. White, "Self-reciprocity transducer calibration in a solid medium," *J. Acoust. Soc. Amer.,* vol. 29, no. 7, p. 834–836, 1957.

[13] J. M. Reid, "Self-reciprocity calibration of echo-ranging transducers," *J. Acoust. Soc. Amer.,* vol. 55, no. 4, p. 862–868, 1974.

[14] M. W. Widener, "The measurement of transducer efficiency using self reciprocity techniques," *," J. Acoust. Soc. Amer.,* vol. 67, no. 3, pp. 1058-1060, 1980.

[15] K. Brendel and G. Ludwig, "Calibration of ultrasonic standard probe transducers," *Acustica,* vol. 36, p. 203–208, 1977.

[16] R. L. Tutwiler, S. Madhavan, and K. V. Mahajan, "Design of test system to characterize very high frequency ultrasound transducer arrays," *Proc. SPIE Med. Imag.,* vol. 3664, p. 182–193, 1999.

[17] A. Caronti, G. Caliano, R. Carotenuto, A. Savoia, M. Pappalardo, E. Cianci, and V. Foglietti, "E. Cianci, and V. Foglietti, "Capacitive micromachined ultrasonic transducer (CMUT) arrays for medical imaging," *Microelectronics Journal,* vol. 37, no. 8, p. 770–777, 2006.

[18] L.W. Schmerr, A. Lopez-Sanchez, R. Huang, "Complete ultrasonic transducer characterization and its use for models and measurements," *Ultrasonics,* vol. 44, p. 753–757, 2006.

[19] D. H. Turnbull and F. S. Foster, "Fabrication and caracterization of tansducer eements in tw-dmensional arrays for medical ultrasound imaging," *IEEE Trans. Ultrason., Ferroelec., Freq. Contr.,* vol. 39, no. 4, pp. 464-475, 1992.

[20] R. C. Chivers, "Time delay spectrometry for ultrasonic transducer characterization," *J. Phys. E: Sci. Instrum.,* vol. 19, no. 10, pp. 834-843, 1986.

[21] B. G. Tomov, S. E. Diederichsen, E. Thomsen, J. A. Jensen, "Characterization of medical ultrasound transducers," in *Proceedings of 2018 IEEE International Ultrasonics Symposium*, 2018.

[22] L. Eghbali, "The impact of defective ultrasound transducers on the evaluation results of ultrasound imaging of blood flow," *MSc Thesis, Royal Institute of Technology , Sweden,* 2014.

[23] M. Vallet, F. Varray, J. Boutet, J. M. Dinten, G. Caliano, A. S. Savoia, and D. Vray, "Quantitative comparison of PZT and CMUT probes for photoacoustic imaging: Experimental validation," *Photoacoustics,* vol. 8, pp. 48-58, 2017.

[24] E. Filoux, J. Mamou, O. Aristizábal, J. A. Ketterling, "Characterization of the spatial resolution of different high-frequency imaging systems using a novel anechoic-sphere phantom," *IEEE Trans Ultrason Ferroelectr Freq Control.,* vol. 58, no. 5, pp. 994-1005, 2011.

[25] C. K. Abbey, N. Q. Nguyen, and M. F. Insana, "Effects of frequency and bandwidth on diagnostic information transfer in ultrasonic B-mode imaging," *IEEE Trans. Ultrason., Ferroelectr., Freq. Control,* vol. 59, no. 6, pp. 1115-1126, June 2012.

[26] J. G. Proakis, Digital Communications, Fourth ed., New York: McGraw-Hill, 2001, p. 234.

[27] R. J. Urick, Principles of Underwater Sound, Third ed., Los Altos: Peninsula Publishing, 1983, pp. 302-306.

[28] W. G. Neubauer, Acoustic Reflection from Surfaces and Shapes, Naval Research Laboratory, 1986, pp. 224-230.

[29] G. S. Kino, Acoustic Waves: Devices, Imaging and Analog Signal Processing, Prentice Hall, 1987, p. 201.

[30] The Engineering Toolbox, [Online]. Available: https://www.engineeringtoolbox.com/sound-speed-solids-d_713.html. [Accessed 4 May 2021].

[31] Y. Kumru, H. Köymen, "Coded Divergent Waves for Fast Ultrasonic Imaging: Optimization and Comparative Performance Analysis," arXiv:2104.10526, 2021.

[32] M. Cikes, L. Tong, G. R. Sutherland, J. D'hooge, "Ultrafast cardiac ultrasound imaging: technical principles, applications, and clinical benefits," *JACC: Cardiovascular Imaging,* vol. 7, no. 8, pp. 812-823, 2014.

[33] M. Lenge, A. Ramalli, P. Tortoli, C. Cachard, H. Liebgott, "Plane wave transverse oscillation for high-frame-rate 2-D vector flow imaging," *IEEE Trans. Ultrason., Ferroelectr., Freq. Control,* vol. 62, no. 12, pp. 2126-2137, Dec. 2015.

[34] S. Bae, J. Jang, M. H. Choi, and T. -K. Song, "In Vivo Evaluation of Plane Wave Imaging for Abdominal Ultrasonography," *Sensors,* vol. 20, no. 19, pp. 1-14, Oct 2020.

[35] G. R. Lockwood, J. R. Talman, S. S. Brunke, "Real-Time 3-D ultrasound imaging using sparse synthetic aperture beamforming," *IEEE Trans. Ultrason., Ferroelectr., Freq. Control,* vol. 45, no. 4, pp. 980-988, July. 1998.

[36] B. Lokesh, A. K. Thittai, "Diverging beam transmit through limited aperture: A method to reduce ultrasound system complexity and yet obtain better image quality at higher frame rates," *Ultrasonics,* vol. 91, pp. 150-160, Jan. 2019.

[37] E. Badescu et al, "Comparison Between Multiline Transmission and Diverging Wave Imaging: Assessment of Image Quality and Motion Estimation Accuracy," *IEEE Trans. Ultrason., Ferroelectr., Freq. Control,* vol. 66, no. 10, pp. 1560-1572, Oct 2019.

[38] E. Roux, F. Varray, L. Petrusca. et al., "Experimental 3-D Ultrasound Imaging with 2-D Sparse Arrays using Focused and Diverging Waves," *Sci Rep.,* pp. 1-12, June 2018.

[39] M. Correia, J. Provost, S. Chatelin, O. Villemain, M. Tanter and M. Pernot, "Ultrafast Harmonic Coherent Compound (UHCC) Imaging for High Frame Rate Echocardiography and Shear-Wave Elastography," *IEEE Trans. Ultrason., Ferroelectr., Freq. Control,* no. 3, pp. 420-431, March 2016.

[40] C. Samson, R. Adamson and J. A. Brown, "Ultrafast Phased-Array Imaging Using Sparse Orthogonal Diverging Waves," *IEEE Trans. Ultrason., Ferroelectr., Freq. Control,* vol. 67, no. 10, pp. 2033-2045, Oct 2020.

[41] J. Kang, D. Go, I. Song and Y. Yoo, "Wide Field-of-View Ultrafast Curved Array Imaging Using Diverging Waves," *IEEE Transactions on Biomedical Engineering,* vol. 67, no. 6, pp. 638-1649, June 2020.

[42] D. Posada, J. Porée, A. Pellissier, B. Chayer, F. Tournoux, G. Cloutier, and D. Garcia, "Staggered Multiple-PRF Ultrafast Color Doppler," *IEEE Transactions on Medical Imaging,* vol. 35, no. 6, pp. 1510-1521, June 2016.

[43] H. Hasegawa, H. Kanai, "High frame rate echocardiography using diverging transmit beams and parallel receive beamforming," *J. Med. Ultrason.,* vol. 38, pp. 129-140, May. 2011.